\documentclass[prd,twocolumn,showpacs,eqsecnum,amsfonts,
amssymb,amsmath]{revtex4}

\newlength{\inda}
\newlength{\indb}

\newcommand{\definition}{:=}
\newcommand{\sa}[1]{{^{#1}\!}}
\newcommand{\YM}[1]{\ensuremath{\mathbb{#1}}}
\newcommand{\ls}{\ell_*}

\begin{document}
\title{Immirzi Ambiguity in the Kinematics of Quantum
General Relativity}
\author{Luis J. Garay and Guillermo A. Mena Marug\'{a}n}
\affiliation{Centro de F\'{\i}sica Miguel A. Catal\'{a}n, I.M.A.F.F.,
C.S.I.C., Serrano 121, 28006 Madrid, Spain}

\begin{abstract}
The Immirzi ambiguity arises in loop quantum gravity when
geometric operators are represented in terms of different
connections that are related by means of an extended Wick
transform. We analyze the action of this transform in gravity
coupled with matter fields and discuss its analogy with the Wick
rotation on which the Thiemann transform between Euclidean and
Lorentzian gravity is based. In addition, we prove that the effect
of this extended Wick transform is equivalent to a constant scale
transformation as far as the symplectic structure and kinematical
constraints are concerned. This equivalence is broken in the
dynamical evolution. Our results are applied to the discussion of
the black hole entropy in the limit of large horizon areas. We
first argue that, since the entropy calculation is performed for
horizons of fixed constant area, one might in principle choose an
Immirzi parameter that depends on this quantity. This would spoil
the linearity with the area in the entropy formula. We then show
that the Immirzi parameter appears as a constant scaling in all
the steps where dynamical information plays a relevant role in the
entropy calculation. This fact, together with the kinematical
equivalence of the Immirzi ambiguity with a change of scale, is
used to preclude the potential non-linearity of the entropy on
physical grounds.
\end{abstract}
\pacs{04.60.Ds, 04.20.Fy, 04.70.Dy} \maketitle

\section{Introduction}

The formulation of general relativity in terms of connection
variables, introduced by Ashtekar \cite{Ash,book}, constitutes one
of the most promising approaches to the quantization of gravity.
In the Ashtekar formalism, the gravitational field is described by
a complex $SU(2)$ connection and a canonically conjugate,
densitized $SU(2)$ soldering form. The shift of emphasis from
geometrodynamics to connection dynamics allows the import of
techniques employed in the quantization of gauge field theories,
providing a common mathematical language for the analysis of
quantum gravity and matter. In addition, the expressions of the
gravitational constraints in Ashtekar variables are extremely
simple, raising renewed hopes for their resolution in the quantum
theory.

The price to be paid is that the Ashtekar connection is complex
for Lorentzian general relativity. This leads to serious technical
and conceptual difficulties, both owing to the lack of a suitable
mathematical machinery to deal with the complex $SU(2)$ group and
because the real part of the Lorentzian connection turns out to
depend on the densitized soldering form, a fact that is
incorporated in the quantization program by imposing the so-called
reality conditions \cite{book,Mat}.

In order to circumvent these problems, essentially two different
avenues have been followed. A possible solution was proposed by
Thiemann, who showed that the Lorentzian and Euclidean sectors of
Ashtekar gravity can be related by an automorphism on the algebra
of functions on phase space \cite{TT,AT}. This automorphism, often
called the Thiemann transform, can be regarded as the composition
of a Wick transform and a complex constant scale transformation
\cite{GM,LG}. It maps the Lorentzian to the Euclidean constraints
and, more importantly, the Ashtekar connection of Lorentzian
general relativity to its Euclidean counterpart, which is real.
Nonetheless, the complications show up when one tries to implement
the Thiemann transform quantum mechanically.

The other possibility was put forward by Barbero. He proved the
existence of a (generalized) canonical transformation that
converts the Lorentzian complex connection into a real connection
\cite{Fer} (which we will call the Ashtekar-Barbero connection
from now on). The only drawback of this change of phase-space
variables is that the expression of the Hamiltonian constraint
loses its original simplicity. But this relative complication is
overwhelmingly compensated for by the availability of the real
$SU(2)$ group as the operationally relevant gauge group.

This real connection formalism has been extensively used for the
quantization of general relativity, mainly in the framework of
loop quantum gravity \cite{loop}. Actually, the Ashtekar-Barbero
connection can easily be generalized to a one-parameter family of
real connections, all of them related by means of canonical
transformations \cite{Fer,Gio}. The associated parameter is
usually called the Immirzi parameter, and we will denote it by
$\gamma$. The remarkable point noticed by Immirzi is that the
physical predictions of the quantum theory depend on $\gamma$.
This is something striking, because the Immirzi parameter
designates just equivalent descriptions of the same phase space.
From a classical point of view, its value does not affect the
physics. Quantum mechanically, however, there exists an ambiguity
in $\gamma$ that appears, e.g., as a multiplicative constant in
the area spectrum \cite{Gio}.

Recently, a radically different alternative to the
Ashte\-kar-Barbero formulation has been suggested which is
apparently free of the Immirzi ambiguity. This alternative
consists in developing a manifestly Lorentz invariant formalism
\cite{Ale}. By retaining the full Lorentz group, one ensures that
the choice of quantization scheme does not result in the
appearance of anomalies, which could cause the Immirzi ambiguity.
In addition, this approach preserves the correct spacetime
interpretation of the gravitational variables. In this sense, it
is worth commenting that the Ashtekar-Barbero connection has been
proved not to transform as the pull-back of a spacetime connection
under diffeomorphisms which are normal to the sections of constant
time \cite{Samuel}. However, the formalism and phase space of this
Lorentz invariant approach are much more intricate than those of
the original Ashtekar-Barbero theory, and further progress is
needed to extract and comprehend its physical predictions.

Since the family of real connections obtained by Immirzi leads to
different quantum results, it is clear that the canonical
transformations that relate these connections cannot be
implemented unitarily \cite{RT}. This obstruction for unitarity is
not well understood. To shed some light on its origin, the Immirzi
ambiguity has been compared with other quantum ambiguities or
anomalies. Rovelli and Thiemann \cite{RT} have tried to construct
a finite dimensional analogue, but their attempt seems to have
been unsuccessful \cite{Sam}. In fact, if one could associate an
independent Immirzi ambiguity with every degree of freedom (a
finite dimensional system), the ambiguity in general relativity
would admit an extension from a constant parameter to two
functions of the spatial position \cite{Samuel}. However, this
hypothetical extension conflicts with diffeomorphism invariance
\cite{Gul}. On the other hand, Corichi and Krasnov have discussed
the possible parallelism between the Immirzi ambiguity and a
factor ambiguity that appears in the electric charge of Maxwell
theory \cite{CK}. But, among standard quantum field theories, the
closest similarity is probably found with the $\theta$ ambiguity
of Yang-Mill theories \cite{GOP}. Unlike what happens in that
case, however, the Immirzi ambiguity does not arise as a
consequence of a multiply connected configuration space
\cite{Sam,Mon}. In this situation, further investigation is
required to clarify the roots and implications of the Immirzi
ambiguity in quantum gravity.

Some proposed interpretations of this ambiguity have been
considered and criticized by Rovelli and Thiemann \cite{RT}, among
them the possibility that the Immirzi parameter amounts to
multiplying the classical action by a constant factor. Although
both ambiguities are not equivalent, a relation between them
should not be discarded. The main reason is that, as pointed out
by Rainer \cite{Martin}, the semiclassical predictions of quantum
gravity may lead to subtle differences between what we call the
Planck length in low-energy physics, $\ell_p$, and what
constitutes the fundamental length scale in the quantum theory,
$\ell_*$. This fundamental length is determined by the overall
factor that multiplies the symplectic structure \cite{Kras} (or,
equivalently, the Poisson brackets). We will return to this issue
later in our work.

In connection with the above comments, it has been argued that the
Immirzi parameter plays simply the role of a scaling of the Planck
length. Since $\gamma$ appears as a global factor in the spectrum
of the area operators \cite{Gio}, the Planck length would be
multiplied by $\sqrt{\gamma}$. However, the scalar constraint
displays a non-homogeneous dependence on $\gamma$ that seems to
conflict with this interpretation \cite{GOP}. One of the aims of
the present paper is to discuss the actual relation between the
Immirzi ambiguity and a constant scale transformation. We will
prove that there indeed exists an equivalence if one restricts all
considerations to the kinematics of the Ashtekar-Barbero
formulation, i.e., if one disregards dynamics. This analysis will
be carried out in gravity with matter fields \cite{Mat}, so that
we can also clarify the extent to which the Immirzi ambiguity is
or is not affected by the introduction of matter (see the
preliminary discussion in Ref. \cite{GOP}).

We will also discuss the implications of our results for one of
the most outstanding predictions of loop quantum gravity: the
entropy formula of a quantum black hole. This entropy is
calculated assuming a horizon with fixed area $\mathcal{A}$ and
adopting a loop quantization with given Immirzi parameter
\cite{QBH,QBH1,IH}. Apparently, therefore, nothing prevents the
value of $\gamma$ from depending on $\mathcal{A}$. This would
destroy the linear dependence on the area in the deduced entropy
\cite{QBH,QBH1}. A way out of this conundrum turns out to be
provided by the kinematical equivalence between a change in
$\gamma$ and a change of scale. As we will show, this equivalence
allows one to regain the Bekenstein-Hawking formula.

The paper is organized as follows. In Sec. \ref{prelim} we
succinctly describe the Ashtekar formalism for gravity coupled
with matter fields. The real Ashtekar-Barbero connection is
introduced in Sec. \ref{AB}, where we also revisit the action of
constant changes of scale and of some suitable extensions of the
Wick and Thiemann transforms. In addition, we analyze the relation
between these extended transforms and the canonical
transformations introduced by Immirzi. Then, we prove in Sec.
\ref{kincons} that the Immirzi ambiguity amounts to a constant
scaling as far as the kinematics of general relativity is
concerned. The physical consequences of this equivalence are
analyzed in Sec. \ref{bhentrop}. In particular, we show that the
effect of the Immirzi ambiguity in the formula of the black hole
entropy can be absorbed into a change of length scale. Our
conclusions are summarized in Sec. \ref{concl}. Finally, an
Appendix is added where we include the expression of the scalar
constraint in the presence of matter fields and study how it is
affected by the Immirzi ambiguity.

\section{Gravity with matter fields}
\label{prelim}

Let us briefly review the Hamiltonian formulation of general
relativity in the presence of matter fields \cite{book,Mat}. We
will consider a matter content consisting of a massive scalar
field $\phi$, massive spin-$1/2$ fields $\xi_A$ and $\bar \eta_A$,
and a Yang-Mills connection $\YM{A}_a$. All these fields are
defined on a certain three-manifold, and we will collectively
denote them as $\{q^k\}$. We call $\{p_k\}$ their respective
canonical momenta $\{\pi_{\phi},\rho^A,\omega^A, \YM{E}^a\}$. By
canonically conjugate we mean variables whose Poisson bracket is
the identity multiplied by $8\pi \ell_*^2$, with
$\ell_*=\sqrt{G\hbar}$. Here, $\hbar$ is the Planck constant
(i.e., the fundamental quantum of action), $G$ is the true Newton
constant that appears in the gravitational action \cite{Kras}, and
we have taken $c=1$. Our notation is very similar to that
introduced in Ref. \cite{LG}. Internal Yang-Mills indices are not
displayed, and spatial indices are denoted with lowercase Latin
letters from the beginning of the alphabet. Capital Latin letters,
on the other hand, designate $SU(2)$ spinors when used as indices.
They are raised and lowered with the alternating tensors
$\epsilon^{AB}$ and $\epsilon_{AB}$ \cite{book}. Whenever they are
not necessary for understanding the formulas, we will also
suppress them.

As for the gravitational part of the phase space, it can be
described by the canonical pair $(a_{aA}^{\;\;\;\;\;B},
i\sqrt{2}\sigma_{\;A}^{a\;\;B})$, where $a_a$ is the (complex)
$SU(2)$ Ashtekar connection and $\sigma^{a}$ is the densitized
soldering form \cite{book}. The Ashtekar connection can be written
as $a_a=\Gamma_a-iV_a$, where $\Gamma_a$ denotes the spin
connection compatible with $\sigma^a$, and $V_a=K_a+iC_a$, with
$K_a$ and $C_a$ being the extrinsic-curvature and fermionic
contributions \cite{book,LG}:
\begin{eqnarray} K_a&=&\frac{K_{ab}\sigma^b}{\sqrt 2 \sigma},
\nonumber\\ C_a^{\;AB}&=&\frac{-i}{4\sqrt 2} (\sigma_{a}^{\;AC}
y_C^{\;\;B}+ \sigma_{a}^{\;BC} y_C^{\;\;A}).
\end{eqnarray}
In this formula, $K_{ab}$ is the extrinsic curvature, $\sigma_{a}$
is the inverse of $\sigma^a$, $\sigma=[\mathrm{det}(-\mathrm{
tr}\{\sigma^a\sigma^b\})]^{1/4}$, and
\begin{equation}
 y_{AB}=  \rho_A\xi_B+ \omega_A\bar\eta_B.
\label{yab}
\end{equation}

For the action proposed by Ashtekar, Romano and Tate
\cite{book,Mat}, the explicit expressions of the kinematical
constraints are the following:
\begin{eqnarray}\label{ashcons}
 \mathbb{G}&=&g^{-1}\YM{D}_a \YM{E}^a,\nonumber\\
 \mathcal{G}_{AB}&=& i\sqrt 2\breve{\mathcal{D}}_a
 \sigma^a_{AB} -y_{(AB)} , \\
 \mathcal{V}_a &=& -i\sqrt 2\mathrm{tr}(\sigma^b
\mathcal{F}_{ab}) -\rho_A \breve{\mathcal{D}}_a \xi^A -\omega_A
\breve{\mathcal{D}}_a\bar\eta^A \nonumber\\
 && -\pi_{\phi}\partial_a\phi -\frac{1}{2}\mathrm{tr} ( \YM{E}^b
\YM{B}_{ab}). \nonumber
\end{eqnarray}
Here, $\mathcal{G}$ and $\mathbb{G}$ are the Gauss constraints
associated with the Ashtekar and Yang-Mills connections (the
latter scaled by the Yang-Mills coupling constant $g^{-1}$ as
compared with that in Ref. \cite{book}), and $\mathcal{V}_a$ is
the vector constraint. The scalar constraint is given in the
Appendix.

We will denote as $\{\chi_l\}$ this set of kinematical constraints
$\{\mathbb{G},\mathcal{G},\mathcal{V}_a\}$. In their expressions,
$\YM{D}_a$ is the derivative operator associated with the
Yang-Mills connection and $\YM{B}_{ab}$ is twice its curvature,
\begin{eqnarray}\label{ymd}
\YM{D}_a \YM{E}^a&= &
\partial_a \YM{E}^a+g [\YM{A}_a,\YM{E}^a],\\ \label{ymb}
\YM{B}_{ab}&=& 2(\partial_a \YM{A}_b-\partial_b
\YM{A}_a+g[\YM{A}_a,\YM{A}_b]).\end{eqnarray} In addition,
$\breve{\mathcal{D}}_a $ is the derivative operator associated
with the Ashtekar connection and $\mathcal{F}_{ab}$ is its
curvature,
\begin{eqnarray}
\breve{\mathcal{D}}_a \xi^A&=&
\partial_a \xi^A-a_{a\;B}^{\;A}\xi^B,\nonumber\\
\mathcal{F}_{ab}&= &\partial_a a_{b} -\partial_b a_{a}
+[a_{a},a_{b}].
\end{eqnarray}

It is worth noting that the only coupling constant on which the
kinematical constraints depend is the Yang-Mills one, $g$. As we
show in the Appendix, the scalar constraint depends not just on
$g$, but also on the fermionic mass $m$ and the scalar field mass
$\mu$. Besides, it contains the cosmological constant $\Lambda$,
if we allow for a cosmological term in the Hamiltonian. We will
generically refer to such parameters as coupling constants and
denote their set as $\kappa\definition \{g,m,\mu,\Lambda\}$.
Finally, the line element can be expressed in terms of the
soldering form, the shift vector $N^a$ and the densitized lapse
function $N$ (with weight equal to $-1$) as
\begin{equation}
ds^2=-\sigma^2 N^2dt^2+h_{ab}(dx^a+N^adt)(dx^b+N^bdt),
\end{equation}
the metric $h_{ab}$ being the inverse of
$h^{ab}=-\sigma^{-2}\mathrm{ tr}(\sigma^a\sigma^b)$.

\section{Ashtekar-Barbero connections}
\label{AB}

The Ashtekar connection is not real for two reasons. First,
because it is a complex linear combination of the spin connection
$\Gamma_a$ and the momenta of the soldering forms $V_a=K_a+iC_a$.
Second, because these momenta $V_a$ are genuinely complex when
fermions are present \cite{book,LG}. With the aim of addressing
the former of these issues, we introduce the connections
\begin{equation} \sa{\gamma} a_a=\Gamma_a-\gamma V_a,
\end{equation}
where $\gamma>0$ is the Immirzi parameter. Since our definition
reproduces that of Ref. \cite{Fer} in the absence of matter, we
will call $\sa{\gamma} a_a$ the Ashtekar-Barbero connections. It
should however be clear that such connections are not real when
fermions are allowed. The canonically conjugate momenta of
$-\sqrt{2}\,\sa{\gamma} a_a$ are
$\sa{\gamma}\sigma^a=\gamma^{-1}\sigma^a$.

Formulations with different values of the Immirzi parameter
$\gamma$ are related by an extended Wick transform $R_\gamma$
\cite{LG,WT} such that
\begin{equation}
\begin{array}{ll}
 R_\gamma\circ \sa{1} a_a=\sa{\gamma} a_a, \qquad & R_\gamma\circ
 q^k=q^k,\\
 R_\gamma\circ\sigma^a=\sigma^a, \qquad & R_\gamma \circ p_k=\gamma
 p_k.
 \end{array}\end{equation}
Note that the action on the gravitational connection can be
rephrased as $R_{\gamma}\circ V_a=\gamma V_a$. Similarly to the
interpretation accepted for the case of the (inverse) Wick
rotation, where $\gamma$ would be equal to $-i$, the simultaneous
multiplication by $\gamma$ of $V_a$ and all the momenta of the
matter fields can be regarded as the consequence of a scaling of
the lapse \cite{LG}. Consequently, the action of $R_{\gamma}$ on
phase space is supplemented with the following transformation laws
on the lapse and shift \cite{WR}:
 \begin{equation}
 R_\gamma\circ N=\gamma^{-1} N,\qquad R_\gamma\circ N^a= N^a.
 \end{equation}
In addition, it seems natural to admit that the transform does not
affect the coupling constants, $R_\gamma\circ\kappa=\kappa$.

This extended Wick transform does not preserve the symplectic
structure. However, as in the case that leads to the Thiemann
transform \cite{LG}, one can complement the extended Wick
transform with a constant scale transformation and construct a
symplectomorphism that relates formulations with different values
of $\gamma$. To show how this can be done, let us first introduce
the constant scale transformations: \begin{equation}
C_{\beta}\circ X=\beta^{D(X)}X.
\end{equation}
Here, $D(X)$ is the dimension of $X$ (a generic field or
parameter). Adopting the convention that the dimensionality of the
line element is carried by the metric and not by the coordinates,
it is possible to show that \cite{LG}
\begin{equation}
\begin{array}{ll}
D(\sa{1} a_a)=0, \quad& D(q^k)= (-1)^{2s_k}s_k,\\
 D(\sigma^a)=2,\quad& D(p_k)= 2-(-1)^{2s_k}s_k, \\
 D(N)=-2, \quad& D(N^a)=0,\\
 D(\Lambda)= -2, \quad&  D(m)=D(\mu)=D(g)=-1.
\end{array}
\end{equation}
In these formulas, $s_k$ denotes the spin of the field $q^k$.

We have now all the ingredients necessary to construct the
extended Thiemann transform $T_{\gamma}$:
\begin{equation}
T_\gamma\definition C_{\gamma^{-1/2}}\circ R_\gamma.
\end{equation}
Its action on phase space is
\begin{eqnarray}
 \sa{\gamma} a_a&\definition&T_\gamma\circ \sa{1} a_a=\Gamma_a-
 \gamma V_a, \nonumber\\
 \sa{\gamma} \sigma^a&\definition&T_\gamma\circ\sigma^a=\gamma^{-1}
 \sigma^a,\nonumber\\ \sa{\gamma} q^k &\definition
 &T_\gamma\circ q^k=\gamma^{-(-1)^{2s_k}
 s_k/2}q^k,\nonumber\\
 \sa{\gamma} p_k&\definition &T_\gamma\circ p_k=\gamma^{(-1)^{2s_k}
 s_k/2}p_k.
\end{eqnarray}
The standard Thiemann transform would be attained with
$\gamma=-i$, except for the fact that our extended transform not
only acts on fields but also on coupling constants. Actually, for
$\gamma=-i$, $T_{\gamma}$ reproduces the modification of the
Thiemann transform proposed in Ref. \cite{AT}, which has a
nontrivial effect on $\kappa$ \cite{LG}. We have
\begin{equation}
T_\gamma\circ\kappa=\gamma^{-D(\kappa)/2}\kappa. \end{equation}

From the above equations, we see that our extended Thiemann
transform preserves the canonical Poisson brackets and canonically
implements the change from one Ashtekar-Barbero connection to
another. Note also that $T_\gamma$ leaves invariant the shift and
densitized lapse,
\begin{equation}
T_\gamma\circ N=N,\qquad T_\gamma\circ N^a=N^a.
\end{equation}
Therefore,
\begin{eqnarray}
\!\!\sa{\gamma} ds^2&\definition &T_\gamma\circ ds^2= -\sigma^2
\gamma^{-3}N^2dt^2 \nonumber\\
&+&\gamma^{-1}h_{ab}(dx^a+N^adt)(dx^b+N^bdt), \label{gds2}
\end{eqnarray}
i.e., it amounts to a constant scale transformation of the line
element plus a change of lapse.

Introducing an index $\alpha$ to denote the different fields
present in the theory (gravitational, scalar, fermionic, and
Yang-Mills fields), the extended Thiemann transform can be further
generalized to
\begin{equation}\label{etfa}
T_{\{\gamma_\alpha\}}=\prod_\alpha T^\alpha_{\gamma_\alpha},
\end{equation}
where $T^\alpha_{\gamma_\alpha}$ is the restriction of
$T_{\gamma_\alpha}$ to the field $\alpha$ (viewed as a canonical
pair of phase-space variables) and its associated coupling
constant $\kappa_\alpha$. In other words, we may allow each field
$\alpha$ to transform according to its own parameter
$\gamma_\alpha$.

It is easy to check that, when acting on phase-space variables,
the transformation $T_{\{\gamma_\alpha\}}$ is generated via
Poisson brackets by \cite{LG}
\begin{equation}
\mathcal{T}_{\{\gamma_\alpha\}}=\frac{i}{2}\sum_\alpha
\ln\gamma_\alpha\, D(q^{\alpha})\int d^3x\, p_\alpha q^\alpha,
\end{equation}
where $\{q^\alpha\}$ denotes the set of variables $\{-\sqrt
2\sigma^a,q^k\}$, their canonically conjugate variables
$\{V_a,p_k\}$ are denoted as $\{p_\alpha\}$, and traces over
$SU(2)$ and Yang-Mills indices are implicitly assumed. The
generator of the standard Thiemann transform is obtained for
$\gamma_\alpha=-i,\ \forall\alpha$ \cite{LG}.

As mentioned above, the connections $\sa{\gamma} a_a$ are still
complex in the presence of fermions, because $V_a$ then has a
non-vanishing imaginary contribution. More precisely, let us use
the Pauli matrices $\tau_{\;A}^{j\;\;B}$ (with $j=1,2,3$) to
express the Ashtekar-Barbero variables in the form \cite{book}
\begin{equation}
\sa{\gamma}\sigma^a=-\frac{i}{\sqrt{2}}\sa{\gamma}\sigma^a_j\tau^j,
\qquad \sa{\gamma} a_a=-\frac{i}{2}\sa{\gamma} a_a^j\tau^j,
\end{equation}
where $\sigma^a_j=\gamma\, \sa{\gamma}\sigma^a_j$ is the
densitized triad. It turns out that the imaginary part of
$V_a^j=i\mathrm{ tr}(V_a\tau^j)$ can be taken equal to
$-i\sigma_a^j\mathrm{tr}(y)/4$ \cite{LG}, which differs from zero
when the system includes fermions. In that case, one can still
recover a real connection by simply replacing $\sa{\gamma} a_a^j$
with its real part, namely, with
\begin{equation}
\sa{\gamma} A_a^j=\sa{\gamma}
a_a^j+\frac{\sa{\gamma}\sigma_a^j}{4}\mathrm{ tr}(\sa{\gamma}y).
\end{equation}
Here, we have employed that $y=\sa{\gamma}y$ and
$\gamma\sigma_a=\sa{\gamma}\sigma_a$. More importantly, it is not
difficult to check that the above change of connections can indeed
be promoted to a canonical transformation by introducing the
following new set of fermionic variables \cite{LG}:
\[
\{\sa{\gamma}\xi',\sa{\gamma}\rho',\sa{\gamma}\bar{\eta}',\sa{\gamma}
\omega'\}=\{\sqrt{\sa{\gamma}\sigma}\,\sa{\gamma}\xi,\sa{\gamma}\rho/
\sqrt{\sa{\gamma}\sigma},\sqrt{\sa{\gamma}\sigma}\,\sa{\gamma}
\bar{\eta},\sa{\gamma}\omega/\sqrt{\sa{\gamma}\sigma}\},
\]
with $\sa{\gamma}\sigma=\sqrt{\mathrm{det}
(\sa{\gamma}\sigma^a_j)}$. In this way, one attains the desired
description of the gravitational field in terms of a real $SU(2)$
connection while preserving the canonical structure on phase
space. Finally, note that from the definition of the new fermionic
fields and the connection $\sa{\gamma}A_a$, the extended Thiemann
transform continues to map canonical variables with $\gamma=1$ to
their counterpart with a generic value of the positive parameter
$\gamma$.

\section{Kinematics and Immirzi ambiguity}
\label{kincons}

The kinematical constraints (\ref{ashcons}) can be written in
terms of the real connections $\sa{\gamma} A_a$ and the rest of
the new canonical variables introduced in the previous section.
After some calculations that involve the compatibility between the
spin connection and the soldering form, the Bianchi identities,
and the relations between Pauli matrices
$(\tau^j\tau^k)_A^{\;\;\;B}=i\epsilon^{jkl}\tau^{l\;\;B}_{\;A}
+\delta^{jk}\delta_A^{\;\,\;B}$ \cite{book}, the constraints
become
\begin{eqnarray}\label{conew}
 \YM{G}&=&\sa{\gamma} g^{-1}\sa{\gamma} \YM{D}_a \sa{\gamma
 }\YM{E}^a\definition\sa{\gamma}\YM{G},\nonumber\\
 \mathcal{G}_{AB}&=& -\sqrt 2\,
\sa{\gamma}\mathcal{D}_a \sa{\gamma}\sigma^a_{AB} -\sa{\gamma}
y'_{(AB)}\definition\sa{\gamma} \mathcal{G}_{AB}, \\
 \mathcal{V}_a &=& \!\sa{\gamma}\mathcal{V}_a
 +\!\frac{i}{4\sqrt 2\gamma}\mathrm{tr}(\sa{\gamma} y')
 \mathrm{tr}(\sa{\gamma} \mathcal{G}\,\sa{\gamma} \sigma_a)
  +\!\frac{i}{\gamma}\mathrm{tr}[\sa{\gamma}\mathcal{G}
  (\Gamma_a\!-\!\sa{\gamma} A_a)],\nonumber
 \end{eqnarray}
where $\sa{\gamma} \mathcal{D}_a$ is the derivative operator
associated with the connection $\sa{\gamma} A_a$, $\sa{\gamma}
F_{ab}$ is its curvature, and
\begin{eqnarray}\label{vecnew}
\sa{\gamma} \mathcal{V}_a&\definition& \sqrt
2\,\mathrm{tr}(\sa{\gamma}\sigma^b \,\sa{\gamma} F_{ab})
-\frac{1}{2}\left(\sa{\gamma} \rho'_A \sa{\gamma} \mathcal{D}_a
\sa{\gamma} \xi^{\prime A}+ \sa{\gamma} \omega'_A \sa{\gamma}
\mathcal{D}_a\sa{\gamma} \bar\eta^{\prime A}\right)
\nonumber\\&+&\frac{1}{2}\left(\sa{\gamma}
\mathcal{D}_a\sa{\gamma} \rho'_A \;\sa{\gamma} \xi^{\prime A}+
\sa{\gamma} \mathcal{D}_a\sa{\gamma} \omega'_A \;\sa{\gamma}
\bar\eta^{\prime A}\right) -\sa{\gamma}
\pi_{\phi}\partial_a\sa{\gamma} \phi \nonumber\\ &-& \frac{1}{2}
\mathrm{tr} ( \sa{\gamma} \YM{E}^b \sa{\gamma} \YM{B}_{ab})
-\mathrm{tr}[\sa{\gamma}\mathcal{G}(\Gamma_a-\sa{\gamma} A_a)].
\end{eqnarray}
In addition, $\sa{\gamma} y'_{(AB)}$ is the counterpart of
expression (\ref{yab}) for the new fermionic variables, and
$\sa{\gamma} \YM{D}_a \sa{\gamma}\YM{E}^a$ and $\sa{\gamma}
\YM{B}_{ab}$ stand for the result of evaluating Eqs. (\ref{ymd})
and (\ref{ymb}), respectively, at the scaled Yang-Mills variables
$(\sa{\gamma}\YM{A}_a,\sa{\gamma}\YM{E}^a)$ employing the new
coupling constant $\sa{\gamma} g\definition C_{\gamma^{-1/2}}\circ
g$.

We hence see that the physical constraints of Lorentz\-ian general
relativity $\{\chi_l\}=\{\mathbb{G},\mathcal{G},\mathcal{V}_a\}$
are equivalent to the new constraints
$\{\sa{\gamma}\chi_l\}=\{\sa{\gamma}\mathbb{G},
\sa{\gamma}\mathcal{G},\sa{\gamma}\mathcal{V}_a\}$. We can
therefore use the latter as the kinematical constraints in the
real connection formulation. Furthermore, note that the functional
dependence of $\sa{\gamma}\chi_l$ on the corresponding set of
canonical variables, specified by the parameter $\gamma$, is the
same for all values of $\gamma$ provided that the Yang-Mills
coupling constant is also scaled with this parameter (according to
the definition of $\sa{\gamma} g$). Actually, this property
continues to hold even when we allow the parameter $\gamma$ to
take different values for each of the fields that are present in
the system, namely, when the canonical variables and coupling
constants denoted with the index $\gamma$ are in fact those
obtained with the set of parameters $\{\gamma_{\alpha}\}$
introduced in Eq. (\ref{etfa}).

It is now straightforward to check that the extended Thiemann
transform $T_{\{\gamma_\alpha\}}$ leaves the kinematical structure
invariant. We have already seen that the transform preserves the
canonical Poisson brackets. Then, we only have to show that it
does not alter the kinematical constraints. By construction,
$T_{\{\gamma_\alpha\}}$ maps the canonical set of variables (and
the coupling constants) with $\gamma=1$ to those with parameters
$\{\gamma_\alpha\}$; so, $T_{\{\gamma_\alpha\}}\circ
\sa{1}\chi_l=\sa{\gamma}\chi_l$ with our notation. But Eqs.
(\ref{conew}) ensure that $\{\sa{\gamma}\chi_l\}$ is equivalent to
$\{\sa{1}\chi_l\}$. The set of constraints $\{\sa{1}\chi_l\}$ is
thus invariant under the extended Thiemann transform. As a
consequence, we can identify the symplectomorphism
$T_{\{\gamma_\alpha\}}$ with the identity transformation at the
kinematical level.

Since, $T_{\{\gamma_\alpha\}}\equiv\openone$ as far as the
kinematical structure is concerned, the action of the transforms
$R_{\{\gamma_\alpha\}}$ and $C_{\{\gamma_\alpha^{1/2}\}}$ can be
regarded as equivalent at this level. It is worth noting that the
latter of these transforms amounts to a constant change of scale
if (and only if) $\gamma_\alpha=\gamma$, $\forall\alpha$. To see
this point clearly, it is convenient to introduce a formalism in
which all fields and coupling constants (that we denote
generically by the symbol $X$) are dimensionless. This can be
accomplished by introducing the universal length scale $\ls$ and
defining $\sa{*} X\definition \ls^{-D(X)}X$, so that
\begin{equation}
C_{\gamma_\alpha^{-1/2}}\circ \sa{*}X =\sa{*}X\quad\text{and}\quad
C_{\gamma_\alpha^{-1/2}}\circ \ls=\gamma_\alpha^{-1/2} \ls.
\end{equation}
Obviously, the scaling transformations of $\ls$ for each kind of
field $\alpha$ are incompatible unless all the parameters
$\gamma_\alpha$ coincide. Then, in the case $\gamma_\alpha=\gamma$
$\forall\alpha$, we see that the action of the extended Wick
transform can indeed be interpreted as a constant scale
transformation if we restrict ourselves to kinematical
considerations:
\begin{equation}\label{iden}
R_\gamma\equiv C_{\gamma^{1/2}}.
\end{equation}

The main reason underlying this result is that both the
constraints $\sa{1}\chi_l$ and the fundamental non-vanishing
Poisson brackets have the same dimension (namely, 2), as well as
the same degree of homogeneity (namely, 1) in the variables
\begin{equation}\{P_{\alpha}\}\definition\{\Gamma_a-\sa{1}A_a,
\pi_{\phi},\sa{1}\rho',\sa{1}\omega',\YM{E}^a\}, \end{equation} up
to terms that vanish because of the Bianchi identities or the
compatibility of the triad with the spin connection. Notice that
the extended Wick transform $R_{\gamma}$ multiplies each of the
elements in $\{P_{\alpha}\}$ by a factor of $\gamma$ while leaving
invariant their canonically conjugate variables
\begin{equation}\{Q^{\alpha}\}\definition\{-\sqrt{2}\sigma^a,
\phi,\sa{1}\xi',\sa{1}\bar{\eta}',\YM{A}_a\}. \end{equation}

We now want to discuss the implications of the equivalence
(\ref{iden}) for the Immirzi ambiguity that arises in loop quantum
gravity at the kinematical level. With this aim, let us consider
an (abstract) operator $\vartheta(\sigma)$ constructed from the
soldering form, its dimension being $D(\vartheta)$. We assume that
this operator is an observable, at least from a kinematical point
of view. Let us call $\mathrm{Sp}[\vartheta(\sigma)]$ and
$\mathrm{Sp}_{\gamma}[\vartheta(\sigma)]$ its spectra in the loop
representations based, respectively, on the real connections
$\sa{1}A_a$ and $\sa{\gamma}A_a$, adopting in both cases as
configuration variables for the matter fields the elements of
$\{Q^{\alpha}\}$ other than the soldering form. Since
$R_\gamma\circ Q^{\alpha}=Q^{\alpha}$ and $R_{\gamma}\circ
\sa{1}A_a=\sa{\gamma}A_a$, the two representations are related via
the extended Wick transform. Using in our kinematical analysis
that $R_\gamma\equiv C_{\gamma^{1/2}}$, one concludes that the
spectrum $\mathrm{Sp}_{\gamma}[\vartheta(\sigma)]$ is the image of
$\mathrm{Sp}[\vartheta(\sigma)]$ under a constant scale
transformation: namely,
\begin{equation}
 \mathrm{Sp}_{\gamma}[\vartheta(\sigma)]=
\mathrm{Sp}[C_{\gamma^{1/2}}\circ\vartheta(\sigma)]=
\gamma^{D(\vartheta)/2}\mathrm{Sp}[\vartheta(\sigma)].
 \label{spect}
\end{equation}
This spectrum is obviously different from
$\mathrm{Sp}[\vartheta(\sigma)]$ if it contains a discrete
component, as happens for instance for the area operator
\cite{area}.

The discrepancy between the spectra of geometric operators in the
loop representations obtained with different connections
$\sa{\gamma}A_a$ is called the Immirzi ambiguity \cite{Gio}. As we
have seen, the ambiguity existing in vacuo persists also in the
presence of matter fields. But much more importantly, our
discussion shows that, owing to the relation $R_\gamma\equiv
C_{\gamma^{1/2}}$, the Immirzi ambiguity can be considered
equivalent to a constant scale transformation at the kinematical
level.

We should note that the presence of matter has introduced a subtle
modification in our analysis with respect to the pure
gravitational situation. As a result of a direct extrapolation of
the conclusions for vacuum gravity, the Immirzi ambiguity is
usually thought to arise in loop representations that are related
by a transform like $R_{\gamma}$, but restricted to act only on
the gravitational field. With the notation of Sec. \ref{AB}, this
would correspond to the replacement of $\gamma$ with a collection
of parameters $\{\gamma_{\alpha}\}$ (one for each field) such that
they all equal the unity except for the gravitational component.
The remarkable point is that the same Immirzi ambiguity appears
when the extended Wick transform acts on all fields with identical
value of $\gamma$. Moreover, as we have proved, the transform can
be consistently interpreted in this case, from a kinematical point
of view, as a constant scale transformation that equally affects
all fields and parameters, not only those associated with the
gravitational sector.

Finally, it should be stressed that the equivalence between the
extended Wick transform (whose implementation results in the
Immirzi ambiguity) and a constant scale transformation is only
valid for kinematical considerations. Indeed, as shown in the
Appendix, the scalar constraint is not invariant under the action
of $T_\gamma$. Therefore, the equivalence does not hold
dynamically and, consequently, the Immirzi ambiguity cannot be
associated with the multiplication of the classical action by a
constant (removable by means of a change of scale) when dynamics
is taken into account \cite{RT,GOP}.

\section{Black hole entropy}
\label{bhentrop}

A particular example of a geometric operator with a discrete
spectrum is that representing the area $\mathcal{A}$ of an
isolated horizon in loop quantum gravity \cite{QBH1,area}. Since
the dimension of the area operator is $2$, Eq.~(\ref{spect}) gives
the relation between the eigenvalues $ \mathcal{A}(J)$ and
$\mathcal{A}_{\gamma}(J)$ in the respective loop representations
with connections $\sa{1}A_a$ and $\sa{\gamma}A_a$:
\begin{equation}
\mathcal{A}_{\gamma}(J)=\gamma \mathcal{A}(J).
\end{equation}
Here, the symbol $J$ labels the different eigenvalues of the
spectrum.

The explicit form of the area spectrum for an isolated horizon,
which is deduced from purely kinematical considerations, is one of
the keystones in the calculation of the statistical entropy $S$ of
a black hole, obtained in the Ashtekar approach by counting
degrees of freedom in the Hilbert space of loop quantum gravity.
An important point in this deduction of the black hole entropy is
that, since the full theory of quantum gravity is not known, the
calculation is carried out by studying only the sector of isolated
horizons with constant area $\mathcal{A}$. This sector is
quantized in a loop representation with a certain Immirzi
parameter $\gamma$. In the limit of a large horizon area, the
resulting entropy is \cite{QBH,QBH1}
\begin{equation}\label{entro}
S=\frac{\gamma_0\mathcal{A}}{4\gamma\ell_*^2},\qquad
\text{with}\quad \gamma_0=\frac{\ln 2}{\pi\sqrt 3}.
\end{equation}
To recover the Bekenstein-Hawking formula
$S=\mathcal{A}/(4\ell_p^2)$ (where $\ell_p$ is the Planck length
of low-energy physics), it is then argued that the Immirzi
parameter must be fixed so that $\gamma\ell_*^2=\gamma_0\ell_p^2$.
In particular, if $\ell_*$ and $\ell_p$ coincide, $\gamma$ must be
chosen equal to $\gamma_0$ in order to reach an acceptable
semiclassical prediction.

There is however a potential loophole in the above line of
reasoning that had remained unperceived so far. Namely, since the
calculation of the black hole entropy is performed for a fixed
value of the horizon area and with a certain choice of the Immirzi
parameter, there seems to be no obstacle to select $\gamma$ in
terms of $\mathcal{A}$, so that $\gamma$ becomes in practice a
function of this quantity. Obviously, this would spoil the linear
relation between the entropy and the area. A way out of this
problem consists in proving that the appearance of the Immirzi
parameter in the entropy formula amounts in fact to a change of
scale. As we will show, imposing that the Planck scale be unique
for all the observers that carry out the measurements will then
eliminate the postulated dependence of $\gamma$ on the area.

Note that, to attain our goal, we only have to demonstrate that
the Immirzi parameter can be absorbed through a constant scaling
in all the steps where the dynamical structure enters the
calculation of the entropy. For all other considerations we simply
have to apply the results of Sec. \ref{kincons}, where we proved
the kinematical equivalence of the Immirzi ambiguity with a change
of scale. Actually, the dynamical structure appears in the entropy
calculation only through the isolated-horizon boundary conditions.
These conditions code dynamical information and intervene in the
analysis inasmuch as they are utilized to prove the following
results \cite{QBH1,IH}:
\begin{enumerate}

\item  The surface terms of the action, necessary for
a well-posed variational problem, correspond to a Chern-Simons
theory whose $U(1)$ connection is
$W_a=\mathrm{tr}(\underline{\Gamma}_a r)$. The underline denotes
the pull-back to the spatial sections of the horizon, which are
topologically $S^2$, and $r=-i r_j\tau^j/\sqrt{2}$ is a fixed
smooth function from the sphere to the Lie algebra of $SU(2)$ with
$\mathrm{tr}(r^2)=-1$. We note that both $r$ and the spin
connection $\Gamma_a$ are invariant under constant scale
transformations.

\item The level $k$ of this $U(1)$ Chern-Simons theory is
$k=\mathcal{A}/(4\pi\gamma\ell_*^2)$, where $\mathcal{A}$ is the
constant area of the horizon. This level determines the overall
factor in the Chern-Simons contribution to the symplectic
structure. Notice that $\gamma$ appears in $k$ exactly as a
scaling of the universal length $\ell_*$.

\item The curvature of the Chern-Simons connection is
$\mathrm{tr}(\sa{\gamma}\underline{\Sigma}_{ab}r)/(2k\ell_*^2)$.
Here, $\Sigma_{ab}\definition \eta_{abc} \sigma^c$ and
$\eta_{abc}$ is the Levi-Civit\`{a} form-density. Since
$\sa{\gamma}\Sigma_{ab}=\gamma^{-1}\Sigma_{ab}$, we see that the
Immirzi parameter appears again as a constant scaling of $\ell_*$.
\end{enumerate}

Remarkably, the isolated-horizon boundary conditions ensure in
addition that there do not exist non-trivial Hamiltonian gauge
transformations at the horizon, so that we do not have to impose
the scalar constraint on it. This means that the dynamical aspects
of the matter content do not affect the physics at the horizon
and, as a consequence, the specific choice of the Immirzi
parameter for the matter fields does not affect the entropy
calculation.

The only issue, purely non-dynamical, remaining in the calculation
of the entropy is counting the physical states whose area
eigenvalue $ \mathcal{A}_{\gamma}(J)$ (in the loop representation
based on $\sa{\gamma}A_a$) lies within the interval
\begin{equation}
\frac{\mathcal{A}}{\gamma\ell_*^2}-\delta\leq
\frac{\mathcal{A}_{\gamma}(J)}{\gamma\ell_*^2}
\leq\frac{\mathcal{A}}{\gamma\ell_*^2}+\delta.
\end{equation}
The counting is made for large $\mathcal{A}/(\gamma\ell_*^2)$ and
assuming that $\delta>4\pi\sqrt 3$ (to ensure the existence of at
least one eigenvalue that corresponds to an even number of spin
insertions). Following the steps in \cite{QBH1}, one then obtains
the entropy formula (\ref{entro}).

As we have already argued, the Immirzi ambiguity is kinematically
equivalent to a constant scale transformation. We have also seen
that all the dynamical arguments involved in the entropy
calculation indicate that $\gamma$ appears as a constant scaling
of the universal length $\ell_*$, even though the scalar
constraint and hence the dynamics in general do not support such
interpretation. Therefore, we can conclude that the appearance of
the Immirzi ambiguity in the entropy formula is equivalent to a
scaling of $\ell_*$.

In our discussion, there seem to exist two length scales which,
for the time being, have been treated as independent
\cite{Martin}. One of them would be the fundamental length scale
$\ell_*$, which appears in front of the action and determines the
Poisson bracket structure \cite{LG,Kras}, and hence the strength
of the quantum gravitational effects. The other would be a
low-energy length scale, which would characterize the low-energy
behavior of quantum gravity, and whose square would provide the
quantum of area. Let us call this length scale the Planck length
and define it as
\begin{equation}
\ell_p=\ell_*\sqrt{\gamma/{\gamma_0}},
\end{equation}
so that
\begin{equation}
S=\frac{\mathcal{A}}{4\ell_p^2}.
\end{equation}
Our definition of $\ell_p$ is feasible because, as we have shown,
the Immirzi ambiguity amounts to a change of scale, at least as
far as the area spectrum and the entropy formula are concerned.
From this point of view, fixing $\gamma$ is equivalent to fixing
the effective value of the low-energy Planck length $\ell_p$ in
terms of the fundamental length $\ell_*$.

Let us remember that, because the whole entropy analysis has been
performed under the assumption that the area of the isolated
horizon is fixed and given a priori, the Immirzi parameter might
in principle be made dependent on $\mathcal{A}$, thus spoiling the
linearity of the relation between entropy and area. Fortunately,
the possible effects of this potential dependence of $\gamma$ on
the area can now be eliminated on the basis of a physical
requirement: when comparing low-energy and large-horizon physics
for different horizons, the comparison must be carried out by
observers that assign the same value to the Planck length
$\ell_p$. Since $\ell_*$ is a universal constant and $\gamma_0$ is
just a numerical factor, this fixes the value of $\gamma$ to be
the same for all observers and, of course, independent of the
area.

\section{Summary and discussion}
\label{concl}

In this paper, we have shown that the Immirzi ambiguity can be
described at a kinematical level in terms of constant scale
transformations. With this aim, we have considered the
Ashtekar-Barbero formulation of general relativity coupled with
fermions, a scalar field, and a Yang-Mills field. In this
framework, the Immirzi ambiguity appears when one calculates the
spectra of geometric operators using loop representations that are
based on different real connections for the gravitational field.
We have shown that these representations can be related via an
extended Wick transform that, in addition to introducing the
Immirzi parameter in the gravitational connection, has also the
effect of multiplying the matter momenta by the same parameter.
This extended Wick transform admits a geometric interpretation as
a scaling of the lapse function, and we have proved that it can be
completed with a constant scale transformation to reach a
symplectomorphism. In a sense, the constructed symplectomorphism
provides an extension of the Thiemann transform that maps the
Lorentzian to the Euclidean formulation of Ashtekar gravity.

Such an extended Thiemann transform has been shown to preserve the
kinematical constraints of the system, so that it can be viewed as
equivalent to the unit transform as far as one disregards the
dynamical evolution. Based on this fact, we have argued that the
Immirzi ambiguity in loop quantum gravity can be understood in
terms of a constant scale transformation for all kinematical
considerations. Indeed, the physical spectra of geometric
operators in loop quantum gravity are affected by the Immirzi
parameter in a way which appropriately depends on the dimension of
the operators. The corresponding scale transformation implies a
change of conformal frame that can be considered responsible for
the quantum ambiguity.

The scalar constraint, on the other hand, is not invariant under
the extended Thiemann transform, and hence the Immirzi ambiguity
cannot be absorbed dynamically into a change of scale. This can
also be rephrased by saying that the four-dimensional line
elements obtained with a constant scale transformation and with a
constant scaling of the lapse do not lead to dynamically
equivalent theories. The breakdown of this equivalence with
respect to the kinematical situation makes us suspicious of the
special role played by time in the Ashtekar-Barbero formulation,
which can be traced back to the time gauge fixing that is
introduced in such a formalism and the consequent loss of a
genuine spacetime interpretation for the gravitational connection
\cite{Gio,Ale, Samuel}.

The Immirzi ambiguity affects one of the most outstanding
predictions of loop quantum gravity, namely, the entropy formula
for isolated horizons. The derivation of this formula involves not
only kinematical but also dynamical processes. However, the
dynamical structure turns out to enter the calculation only coded
in the isolated-horizon boundary conditions. It is worth
commenting that an analogous conclusion has been recently reached
by Padmanabhan from a completely independent point of view
\cite{Pad}. He has deduced the entropy of a spherically symmetric
spacetime with a horizon by studying the partition function of a
canonical ensemble with fixed temperature on that horizon. This
analysis does not assume that the spacetimes in the ensemble are
solutions to the Einstein equations; moreover, it is seen that the
result depends only on the form of the metric near the boundary
supplied by the horizon.

Returning to our study of the loop approach, we have shown that,
whenever the Immirzi parameter appears in the calculation of the
black hole entropy through the conditions on the horizon,
remarkably, it behaves in fact as though it came from a constant
scale transformation. The remaining arguments that lead to the
entropy formula are strictly kinematical. It then follows from our
discussion that the Immirzi ambiguity in the relation between
entropy and area can be understood as a conformal ambiguity in the
length scale employed to measure large horizon areas and
low-energy processes in general. This is important because the
entropy calculation is performed for isolated horizons with a
constant area that is given a priori. Therefore, the choice of
Immirzi parameter in the loop quantization might in principle be
made dependent on this area, and this would ruin the possibility
of deducing the Bekenstein-Hawking formula. However, this
potential dependence on the area of the Immirzi parameter
disappears if we insist that the results for different large
isolated horizons be compared by observers which agree on the
value of the length scale that controls the semiclassical
gravitational effects from the low-energy point of view, i.e.,
Planck length, because the Immirzi parameter can be absorbed into
this length scale.

\begin{acknowledgments}

G.A.M.M. is very thankful to J.F. Barbero G. for enlightening
discussions. This work was supported by funds provided by the
Spanish Ministry of Science and Technology under the Research
Project No. BFM2001-0213.

\end{acknowledgments}

\appendix

\section{Scalar constraint}

In this appendix, we present the expressions of the scalar
constraint in terms of the Ashtekar connection and of the real
connections $\sa{\gamma}A_a$. In the Ashtekar formulation of
Lorentzian gravity with matter fields, the scalar constraint is
\cite{book,Mat,LG}:
\begin{eqnarray}
 \mathcal{S}&=& -\mathrm{tr}
 (\sigma ^a \sigma^b \mathcal{F}_{ab})
  +i m(\sigma^2\xi^A\bar\eta_A-\rho^A\omega_A) \nonumber\\
 &+&i\sqrt 2  \sigma_{\;A}^{a\;\;B}(\rho_B\breve{\mathcal{D}}_a
 \xi^A+\omega_B\breve{\mathcal{D}}_a\bar\eta^A)+\sigma^2
 \Lambda\nonumber\\
 &+&\frac{\pi_{\phi}^2}{16\pi}
 +4\pi \sigma^2\mu^2\phi^2 -4\pi\mathrm{tr}(\sigma^a\sigma^b)
 \partial_a\phi\partial_b\phi \nonumber\\
 &+&\frac{1}{8\sigma^2}\mathrm{tr}
 (\sigma^a\sigma^c ) \mathrm{tr}(\sigma^b\sigma^d )\mathrm{tr}
 (\YM{E}_{ab}\YM{E}_{cd}+\YM{B}_{ab}\YM{B}_{cd}),
 \nonumber
 \end{eqnarray}
where we have employed the notation of Sec. \ref{prelim}, $m$ and
$\mu$ denote the masses of the fermionic and scalar fields,
$\Lambda$ is the cosmological constant, and
$\YM{E}_{ab}=\eta_{abc} \YM{E}^c$, with $\eta_{abc}$ being the
Levi-Civit\`{a} form-density.

This constraint leads to the Einstein-Cartan theory, which is
quartic in the fermionic variables. Nevertheless, one can attain
the Einstein-Dirac theory, quadratic in fermionic fields, by
simply adding to $\mathcal{S}$ the term \cite{book,LG}:
\[
\mathcal{S}_{\rm f}=-\frac{3}{16} \big( y_A^{\;\;A} y_B^{\;\;B}+
 y_{AB}  y^{AB} + y_{AB} y^{BA}\big).
\label{sf}
\]

In terms of the real connection $\sa{\gamma}A_a$, the scaled
soldering form $\sa{\gamma}\sigma^a$, and the canonical set of
matter variables specified by the parameter $\gamma$ (here, we
concentrate on the case of interest $\gamma_\alpha=\gamma$
$\forall \alpha$), the scalar constraint can be written as
follows:
\begin{eqnarray}\label{sco}
 \mathcal{S}&=&\gamma^2\;\sa{\gamma}\,\mathcal{S}
 +\frac{i}{\sqrt 2}\gamma(\gamma-1)\sa{\gamma}Y-
 (\gamma^2-1)\sa{\gamma}Z\nonumber \\
 &-&\frac{i}{\sqrt 2}\gamma\; D_a\mathrm{tr}(\sa{\gamma}
 \mathcal{G}\,\sa{\gamma}\sigma^a),\end{eqnarray}
where $D_a$ is the derivative operator compatible with the triad,
obtained with the spin connection $\Gamma_a$, and
\begin{eqnarray}
 \sa{\gamma}\mathcal{S}&\definition& \!-\mathrm{tr}
 (\sa{\gamma}\sigma ^a \,\sa{\gamma}\sigma^b\,\sa{\gamma}{F}_{ab})
  \!+\mathrm{tr}(\sa{\gamma}\sigma^a\,\sa{\gamma}\sigma^b[\Gamma_a
 \!-\!\sa{\gamma}A_a,\Gamma_b\!-\!\sa{\gamma}A_b]) \nonumber\\
 &+&\sa{\gamma}\sigma^2\,\sa{\gamma}\Lambda+i \sa{\gamma}m
 \sa{\gamma}\sigma \sa{\gamma}\xi^{\prime A}\,\sa{\gamma}\bar\eta'_A+
 \frac{1}{\sqrt 2}\mathrm{tr}(\sa{\gamma}\sigma D_a\sa{\gamma}y')
 \nonumber\\
 &+&4\pi\,\sa{\gamma}\sigma^2\,\sa{\gamma}\mu^2\,\sa{\gamma}\phi^2-
 4\pi\mathrm{tr}(\sa{\gamma}\sigma^a\,\sa{\gamma}\sigma^b)
 \partial_a\sa{\gamma}\phi\,\partial_b\sa{\gamma}\phi
 \nonumber\\
 &+&\frac{1}{8\,\sa{\gamma}\sigma^2}\mathrm{tr}(\sa{\gamma}\sigma^a
 \,\sa{\gamma}\sigma^c )\mathrm{tr}(\sa{\gamma}\sigma^b\,
 \sa{\gamma}\sigma^d )\mathrm{tr}(\sa{\gamma}\YM{B}_{ab}
 \sa{\gamma}\YM{B}_{cd})\nonumber\\
 &+&\sa{\gamma}Z-\frac{i}{\sqrt 2}\sa{\gamma}Y+\frac{1}
 {\sqrt 2}D_a\mathrm{tr}(\sa{\gamma}\mathcal{G}\,
 \sa{\gamma}\sigma^a),\nonumber\end{eqnarray}
\begin{eqnarray}
\sa{\gamma}Y&\definition& \!\mathrm{tr}
 \big(\sa{\gamma}\sigma^a\{D_a\!\sa{\gamma}\rho'\,\sa{\gamma}
 \xi'\!\!+\!D_a\!\sa{\gamma}\omega'\,\sa{\gamma}\bar\eta'
 \!-\!\sa{\gamma}\rho'{D}_a\!\sa{\gamma}\xi'
 \!-\!\sa{\gamma}\omega'{D}_a\!\sa{\gamma}\bar\eta'\}\big),
 \nonumber\\
 \sa{\gamma}Z&\definition&
 \mathrm{tr}(\sa{\gamma}\sigma^a\,\sa{\gamma}\sigma^b
 [\Gamma_a-\sa{\gamma}A_a,\Gamma_b-\sa{\gamma}A_b])
 +\frac{3}{16}[\mathrm{tr}(\sa{\gamma}y')]^2 \nonumber\\
 &-&\!\!i \sa{\gamma}m\sa{\gamma}\sigma\sa{\gamma}\rho^{\prime A}\,
 \sa{\gamma}\omega'_A-\frac{1}{\sqrt 2}\mathrm{tr}
 (\sa{\gamma}y'\tau^k)\mathrm{tr}(\sa{\gamma}\sigma^a
 \tau^k\{\Gamma_a-\sa{\gamma}A_a\})\nonumber\\
  &+&\frac{\sa{\gamma}\pi_{\phi}^2}{16\pi}+
  \frac{1}{8\,\sa{\gamma}\sigma^2}\mathrm{tr}(\sa{\gamma}
  \sigma^a\,\sa{\gamma}\sigma^c )\mathrm{tr}(\sa{\gamma}
  \sigma^b\,\sa{\gamma}\sigma^d )\mathrm{tr}
 (\sa{\gamma}\YM{E}_{ab}\sa{\gamma}\YM{E}_{cd}).
\nonumber\end{eqnarray} We have used the notation introduced in
Secs. \ref{AB} and \ref{kincons}, and the scaled coupling
constants are $\sa{\gamma}\kappa\definition
\gamma^{-D(\kappa)/2}\kappa$. $D(\kappa)$ is equal to $-1$ except
for $\Lambda$, whose dimension is $-2$.

Taking $\gamma=1$, we see from the above expressions that
$\sa{1}\mathcal{S}=\mathcal{S}$ modulo the gravitational Gauss
constraint. We can hence use $\sa{1}\mathcal{S}$ as the scalar
constraint in the real connection formulation with $\gamma=1$. On
the other hand, recalling that the different sets of canonical
phase-space variables and coupling constants parametrized by
$\gamma$ are related by the extended Thiemann transform $T_\gamma$
and using our definition of $\sa{\gamma}\mathcal{S}$, we
straightforwardly obtain that
$T_\gamma\circ\sa{1}\mathcal{S}=\sa{\gamma}\mathcal{S}$. However,
$\sa{\gamma}\mathcal{S}$ and $\sa{1}\mathcal{S}$ differ when
$\gamma\neq 1$ even modulo the kinematical constraints, as can be
easily checked from Eq. (\ref{sco}). As a consequence, the
dynamical structure is not invariant under the extended Thiemann
transform. Therefore, the kinematical equivalence
$R_{\gamma}\equiv C_{\gamma^{1/2}}$ is not maintained when
dynamics is taken into account.


\begin{thebibliography}{30}
\bibitem{WT} This extended Wick transform should not be mistaken for
the Thiemann transform \cite{TT,AT}, which is sometimes referred
to as the generalized Wick transform.
\bibitem{WR} Our notation does not make distinctions between
fields that depend on the different time coordinates
corresponding, respectively, to the original and scaled lapses $N$
and $R_{\gamma}\circ N$. This differs from the notation adopted in
Ref. \cite{LG}, where Euclidean and Lorentzian fields were
distinguished for the case of the Wick rotation.
\end{thebibliography}
\end{document}